\documentclass[cernpreprint,texlive=2011,txfonts,UKenglish]{latex/atlasdoc}

\usepackage[backend=biber]{latex/atlaspackage}
\usepackage{latex/atlasphysics}
\usepackage{latex/atlasbiblatex}
\usepackage{subfig}

\newcommand{\AtlasCoordFootnote}{
ATLAS uses a right-handed coordinate system with its origin at the nominal interaction point (IP)
in the centre of the detector and the $z$-axis along the beam pipe.
The $x$-axis points from the IP to the centre of the LHC ring,
and the $y$-axis points upwards.
Cylindrical coordinates $(r,\phi)$ are used in the transverse plane,
$\phi$ being the azimuthal angle around the $z$-axis.
The pseudorapidity is defined in terms of the polar angle $\theta$ as $\eta = -\ln \tan(\theta/2)$.
Angular distance is measured in units of $\Delta R \equiv \sqrt{(\Delta\eta)^{2} + (\Delta\phi)^{2}}$.}

\AtlasTitle{Search for resonances in the mass distribution of jet pairs with one or two jets identified as $\boldmath b$-jets in proton--proton collisions at $\sqrt{s}=13$~TeV with the ATLAS detector}

\AtlasJournal{\PLB}

\PreprintIdNumber{CERN-EP-2016-066}

\AtlasAbstract{
Searches for high-mass resonances in the dijet invariant mass spectrum with one or two jets identified as $b$-jets are performed using an integrated luminosity of $3.2~$fb$^{-1}$ of proton--proton collisions with a centre-of-mass  energy of $\sqrt{s}=13$~TeV recorded by the ATLAS detector at the Large Hadron Collider.
No evidence of anomalous phenomena is observed in the data, which are used to exclude, at 95\%~credibility level, excited $b^{*}$ quarks with masses from 1.1~\TeV~to 2.1~\TeV~and leptophobic $Z'$ bosons with masses from 1.1~\TeV~to 1.5~\TeV. Contributions of a Gaussian signal shape with effective cross sections ranging from approximately 0.4 to 0.001~pb are also excluded in the mass range 1.5--5.0~\TeV.
}

\begin{document}

\maketitle

\clearpage

\section{Introduction}
\label{sec:intro}
Many extensions to the Standard Model (SM) predict the existence of new massive particles that couple to quarks or gluons. If produced  in proton--proton ($pp$) collisions at the Large Hadron Collider (LHC), these new beyond-the-SM (BSM) particles could decay into quarks ($q$) or gluons ($g$), creating resonant excesses in the two-jet (dijet) invariant mass distributions~\cite{dijetreview1,dijetUA1,dijetUA2,dijetCDF,ATLAS:2015nsi,CERN-PH-EP-2015-317}.
If the new particle couples to the $b$-quark and decays into $b\bar{b}$, $bq$ or $bg$ pairs, a dedicated search for dijet resonances with one or both jets
identified as originating from a $b$-quark (``$b$-jet'') could greatly increase the signal sensitivity.

Prior resonance searches in dijet events containing $b$-jets were performed by the CDF~\cite{Abe:1998uz} and CMS~\cite{cmsdibjet1,cmsdibjet2} experiments, probing the mass ranges 200--750$\GeV$ and 1--4$\TeV$ respectively. Excited heavy-flavour quarks have been investigated in alternative decay modes as well~\cite{Aad:2015voa}. No BSM phenomena have been observed yet.
The increase in centre-of-mass energy of the $pp$ collisions at the LHC from $\sqrt{s}=7$ and $8\TeV$ to $13\TeV$ provides a new energy regime in which to search for such a heavy resonance. This is particularly true for heavy states coupling to $b$-quarks from the proton sea, when compared to states produced by valence quarks.
The parton luminosity to create a $2 \TeV$ object increases by an additional factor of 2--3 for $b\bar{b}$ and $bg$ over $q\bar{q}$ and $qg$ pairs, when increasing the centre-of-mass energy from $8\TeV$ to $13\TeV$. The total production rate for dijet BSM signals
can become large enough to allow a good signal sensitivity even with a relatively small data sample.
In this paper the search for a new narrow resonance decaying to $b$-quarks with the ATLAS detector, using
$3.2~\ifb$ integrated luminosity of proton--proton collisions at $\sqrt{s}=13\TeV$, is reported. The mass range 1.1--5.0$\TeV$ is probed.

\begin{figure}[htbp]
\center
\subfloat{\includegraphics[width=0.45\linewidth]{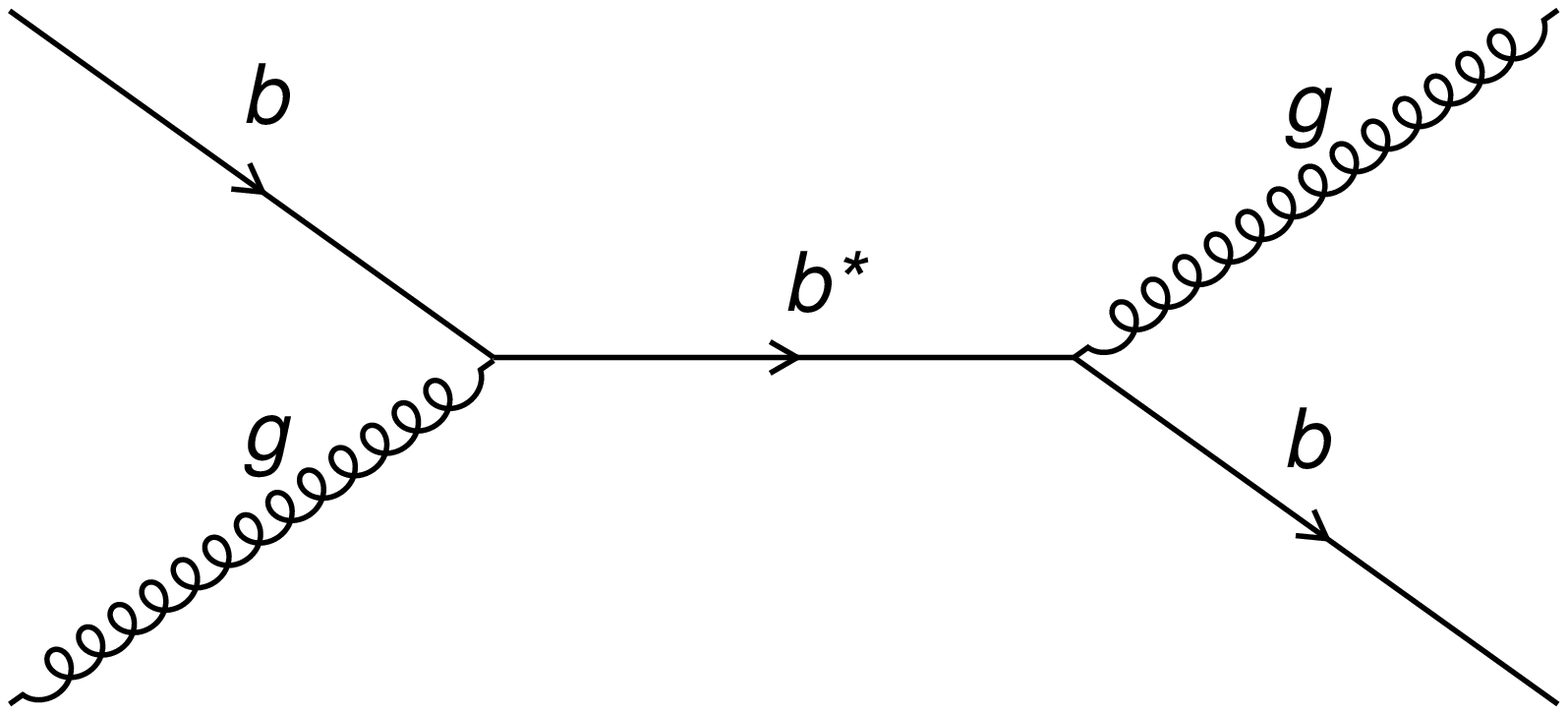}}
\subfloat{\includegraphics[width=0.45\linewidth]{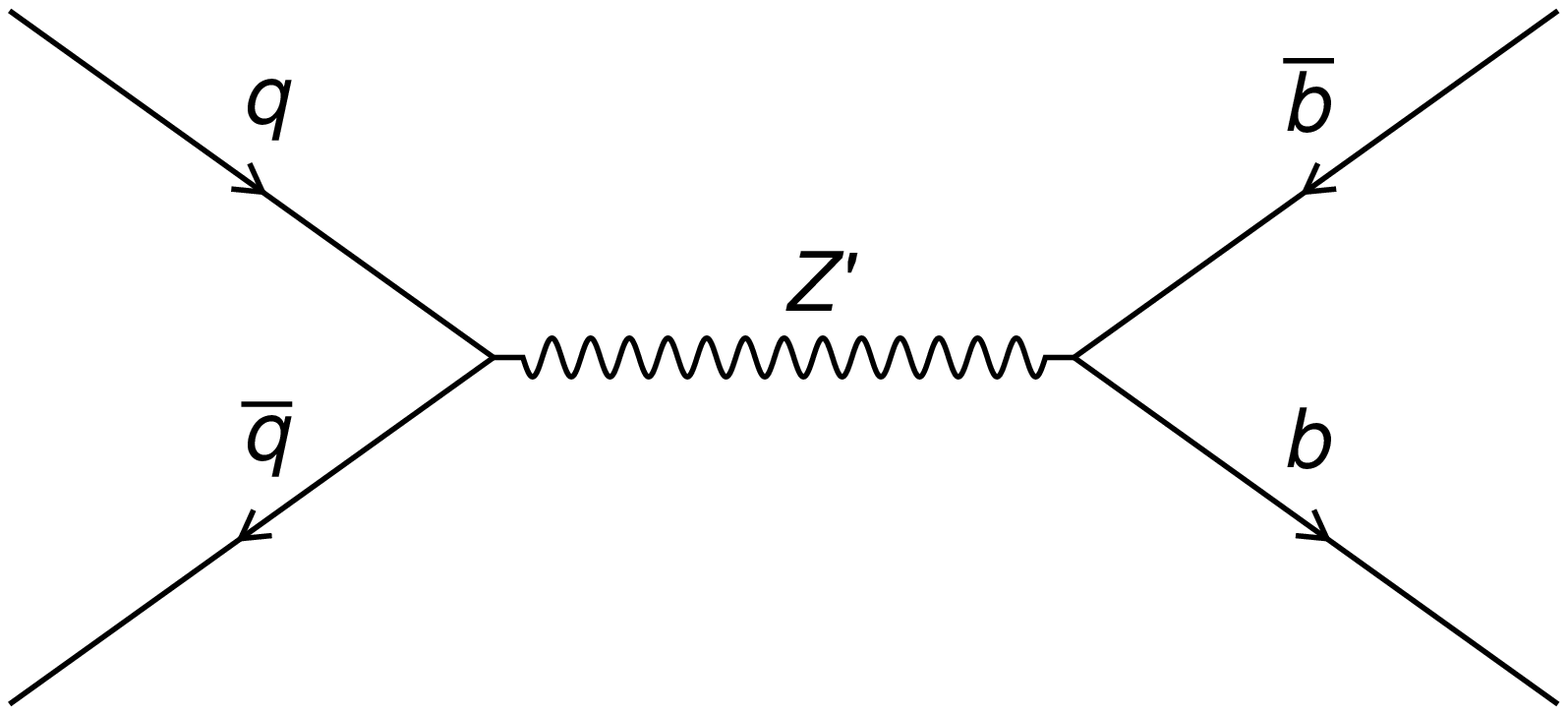}}
\caption{Leading-order Feynman diagrams for the two processes considered: $gb\to b^*\to bg$ and $q\bar{q}\to Z'\to b\bar{b}$.}
\label{fig:Feynman}
\end{figure}

The results are interpreted in the context of two benchmark processes shown in Figure~\ref{fig:Feynman}: an excited heavy-flavour quark $b^*$ and a new gauge boson $Z'$.
Excited quarks are a consequence of quark compositeness models that were  proposed to explain the generational structure and mass hierarchy of quarks~\cite{Baur:1987ga,PhysRevD.42.815}.
The $Z'$ boson arises in many extensions to the SM with an additional $U(1)$ group.
Two $Z'$ models are considered, one with SM-like fermion couplings in the Sequential Standard Model (SSM) and a leptophobic $Z'$ model~\cite{Langacker:2008yv,Chiang:2014}.
All benchmark model decays are expected to result in a narrow resonance superimposed on a smoothly falling dijet invariant mass distribution.
This search divides the events into samples with one or two jets identified as $b$-jets to enhance the signal sensitivity to the benchmark models $b^*\to bg$ and $Z'\to b\bar{b}$.
In addition, the results are interpreted in the context of possible Gaussian-shaped signal contributions to the dijet invariant mass spectra where one or both jets are identified as b-jets.  The results, presented in terms of the the cross section times acceptance times branching ratio $(\sigma\times A\times\rm BR)$, are quoted for contributions with widths of up to $15\%$ of the resonance mass.

\section{The ATLAS detector}
\label{sec:atlas}

The ATLAS experiment~\cite{PERF-2007-01} at the LHC is a multi-purpose particle detector with a forward-backward symmetric cylindrical geometry and a near $4\pi$ coverage in solid angle.\footnote{\AtlasCoordFootnote}
It consists of an inner tracking detector surrounded by a thin superconducting solenoid providing a \SI{2}{\tesla} axial magnetic field, electromagnetic and hadronic calorimeters, and a muon spectrometer.
The inner tracking detector covers the pseudorapidity range $|\eta| < 2.5$. It consists of, in ascending order of radius from the beam-line, silicon pixel, silicon microstrip, and transition radiation tracking detectors. The pixel detectors are crucial for $b$-jet identification. For the second LHC data-taking period, a new inner pixel layer, the Insertable B-Layer (IBL)~\cite{ibl,iblAdd}, was added at a mean sensor radius of 3.2~cm from the beam-line.
Lead/liquid-argon (LAr) sampling calorimeters provide electromagnetic (EM) energy measurements with high granularity.
A hadron (steel/scintillator-tile) calorimeter covers the central pseudorapidity range ($|\eta| < 1.7$).
The end-cap and forward regions are instrumented with LAr calorimeters for EM and hadronic energy measurements up to $|\eta| = 4.9$.
The first-level trigger is implemented in hardware and uses a subset of the detector information to reduce the input rate from the nominal LHC collision rate to an acceptance rate of \SI{100}{\kHz}. This is followed by a software-based trigger that reduces the rate of events recorded to \SI{1}{\kHz}.

\section{Data and simulated event samples}
\label{sec:dataandmc}
The data used in this analysis were collected by the ATLAS detector in $pp$ collisions at the LHC with a centre-of-mass energy of $13\TeV$ during 2015. Events were recorded using a jet-based trigger requiring at least one jet with a transverse momentum $p_\mathrm{T}$ of at least $360\GeV$.
The full dataset corresponds to an integrated luminosity of $3.2~\ifb$ with an associated uncertainty of 5\% after applying quality criteria to the data.
The measurement of the integrated luminosity is derived, following a methodology similar to that detailed in Ref.~\cite{DAPR-2011-01}, from a calibration of the luminosity scale using a pair of $x$--$y$ beam-separation scans.

Monte Carlo (MC) simulated event samples are used to model the expected signals and study the composition
of SM background processes.
The QCD dijet process is simulated with {\sc Pythia8}~\cite{pythia8} using the A14 tuned parameter set~\cite{a14tune} for the modelling
of the parton shower, hadronization and underlying event. The leading-order (LO) parton distribution function (PDF) set NNPDF2.3~\cite{nnpdf}
is used for the generation of events. The renormalization and
factorization scales are set to the average transverse momentum $p_\mathrm{T}$ of the two leading jets.
The \textsc{EvtGen} decay package~\cite{evtgen} is used for bottom and charm hadron decays.

The three signal samples are generated with {\sc Pythia8} using the A14 set of tuned parameters and the NNPDF2.3 PDF set.
For the $b^{*}$ model, the compositeness scale is set to the excited-quark mass and 85\%~of decays are to $bg$. The remaining decay modes are to a SM gauge boson ($Z$ boson,  $W$ boson or photon) and a $b$-quark.
In the SSM $Z'$ model, the $Z'$ boson has the same couplings to SM fermions as the SM $Z$ boson and the bottom quark decay branching ratio $\rm{BR}$$(Z'\rightarrow b\bar{b})$ is $13.8\%$. The leptophobic $Z'$ model differs by having vanishing couplings to leptons. The corresponding value of $\rm{BR}$$(Z'\rightarrow b\bar{b})$ is $18.9\%$. For both, only decays to $b$-quark pairs are simulated.
The intrinsic decay width is $\Gamma\sim 0.6\%$ of the resonance mass for the $b^{*}$ model and $\Gamma\sim 3\%$ of the mass for the SSM $Z'$ boson.

The generated samples are processed with the ATLAS detector simulation~\cite{atlassim}, which is based on the {\sc GEANT4} package~\cite{geant4}. To account for additional $pp$ interactions from the same or close-by bunch crossings, a number of minimum-bias interactions generated using {\sc Pythia8} and the MSTW2008LO PDF~\cite{mstw} set are
superimposed onto the hard scattering events. The MC samples are re-weighted to match the collisions per bunch crossing observed in the data.

\section{Event reconstruction and  selection}
\label{sec:selection}

Jets are reconstructed from noise-suppressed topological clusters~\cite{Aad:2016upy} of energy deposited in the calorimeters using the anti-$k_{t}$ algorithm~\cite{antikt} with a radius parameter of 0.4. Jet energies and directions are corrected by the jet calibrations derived from $\sqrt{s}=13\TeV$ simulation, and $pp$ collision data taken at $\sqrt{s}=8\TeV$ and $\sqrt{s}=13\TeV$, as described in Ref.~\cite{ATL-PHYS-PUB-2015-015}.  Jets are required to have $p_\mathrm{T}>50\GeV$.
Events where any of the three leading jets with $p_\mathrm{T}>50\GeV$ is compatible with non-collision background or calorimeter noise are removed.
Events are preselected in the same way as in the dijet analysis of Ref.~\cite{ATLAS:2015nsi}, requiring that the $p_\mathrm{T}$ of the leading jet is greater than $440\GeV$ to ensure full trigger efficiency. An additional requirement is placed on the jet pseudorapidity, $|\eta|<2.4$, to ensure tracker coverage for $b$-jet identification.
The analysis is performed in an unbiased dijet mass range of $m_{\mathrm{jj}} > 1.1\TeV$.
To reduce the background from QCD multijet processes and enhance $s$-channel processes, the rapidity difference $y^* = (y_1 - y_2)/2$ between the two leading jets is required to be $|y^{*}|<0.6$. Here $y_1$ and $y_2$ are the rapidities of the leading and sub-leading jet respectively.

To identify jets originating from $b$-hadrons ($b$-tagging) a multivariate algorithm
that combines information about the impact parameters of inner detector tracks associated with the jet, the presence of displaced secondary vertices, and the reconstructed flight paths of $b$- and $c$-hadrons associated with the jet~\cite{Aad:2015ydr,btagging} is employed. The $b$-tagging working point with $85\%$ efficiency, as determined when integrating over all jets in a simulated sample of $t\bar{t}$ events, is chosen because it gives the highest signal sensitivity. As the average jet energies in this analysis are larger than in $t\bar{t}$ events and the $b$-tagging efficiency drops with jet $p_\mathrm{T}$, the per-jet efficiencies are below $85\%$ and are roughly $50\%$ for jets with a $p_\mathrm{T}$ of $1\TeV$.

\begin{figure}[htbp]
  \center
\includegraphics[width=0.65\linewidth]{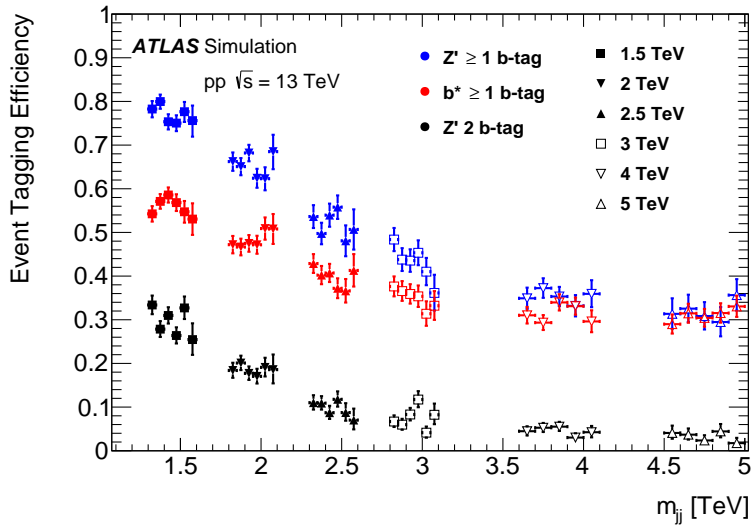}
  \caption{The per-event $b$-tagging efficiencies after the event selection as a function of the reconstructed invariant mass for simulated samples with six different $b^*$ and $Z'$ resonance masses.
  \label{fig:btageff}}
\end{figure}

The $b$-jet identification algorithm is applied to the two leading jets, and events are categorized as inclusive, single $b$-tagged ``1b'' or double $b$-tagged ``2b'', in order to enhance the sensitivity of different signal compositions. The ``1b'' category is defined inclusively, including events from the ``2b'' category.

The per-event $b$-tagging efficiencies as functions of the reconstructed invariant mass are shown in Figure~\ref{fig:btageff}.
Efficiencies are for benchmark models with different $b^*$ and $Z'$ resonance masses, after the event selection is applied.
The tagging efficiency for $Z'$ events in the inclusive ``1b'' category is higher than for $b^*$ events because this process has more $b$-quarks in the final state.
At high mass, the gluon from the decay of the $b^*$ has a higher probability to produce a $b\bar{b}$-pair, which causes the event tagging efficiency to be comparable for the $Z'$ and $b^*$. The tagging efficiency in the ``2b'' category is about 2.5~times lower at low mass and a factor~10 lower at high mass compared to the inclusive ``1b'' category for the same $Z'$ events. The average light-flavour jet rejection factor for jets passing the kinematic selection is approximately 30 for jet transverse momenta up to $\sim 1\TeV$.

Correction factors are applied to the simulated event samples to compensate for differences between data and simulation in b-tagging efficiencies and mis-identification rates. These corrections were derived from comparisons of samples of $b$-quark-enriched events in data and simulation~\cite{ATL-PHYS-PUB-2015-039}.
The average combined signal acceptance and efficiency is around 20\% for the $b^*$ benchmark in the ``1b'' category and drops with increasing mass from 9\% at $1.5~\TeV$ to 2\% at $5.0~\TeV$ for the $Z'$ signals for the ``2b'' category.

\section{Dijet mass spectrum}
\label{sec:dijetmass}

The dijet mass spectrum is predominantly composed of jets arising from QCD interactions. Figure~\ref{fig:mjj} shows the comparison between data and {\sc Pythia8} multijet MC simulation. The simulated distributions are normalized to the number of events observed in the data in each category separately. The bin widths are chosen to approximate the $m_{\mathrm{jj}}$ resolution as derived from simulated QCD processes, which range from $3\%$ at $1.0\TeV$ to $2\%$ at $5.0\TeV$. Good agreement between the shapes of the {\sc Pythia8} multijet predictions and the data is found. The inclusive distribution, not restricted in the inner tracking detector acceptance, was analysed in Ref.~\cite{ATLAS:2015nsi}.

\begin{figure}[htbp]
  \center
  \includegraphics[width=0.95\linewidth]{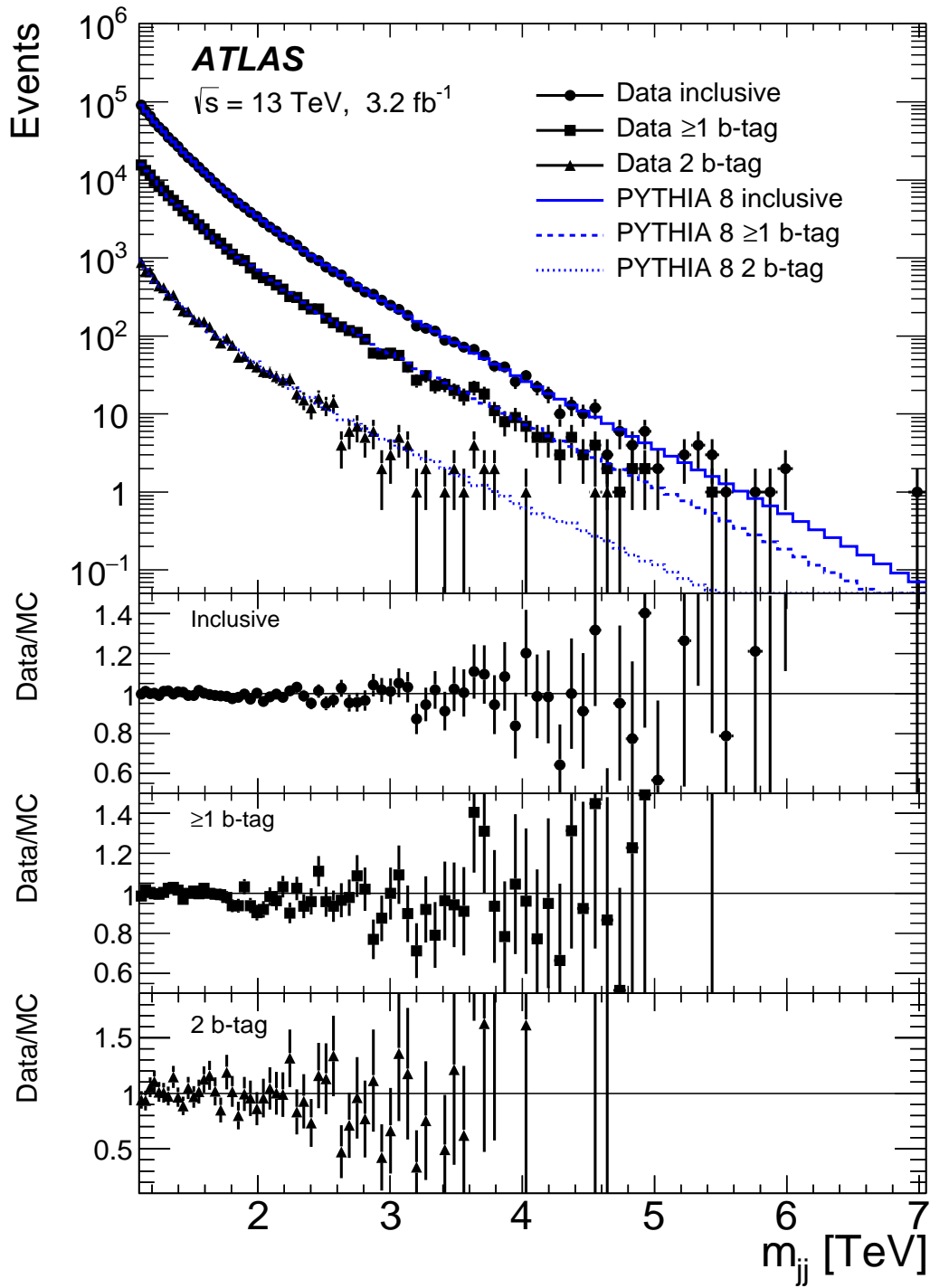}
  \caption{The invariant mass distribution of the inclusive dijet (dots), ``1b'' (squares) and ``2b'' (triangles) categories in data. The inclusive distribution is similar to Ref.~\cite{ATLAS:2015nsi}, but an additional requirement is placed on the jet pseudorapidity, $|\eta|<2.4$. The MC distributions are normalized to the data in the three categories separately: a solid line for inclusive dijets, a dashed line for ``1b'' and a small dashed line for ``2b'' categories. The lower panels show the ratio between data and MC simulation for all three categories.}
  \label{fig:mjj}
\end{figure}

The dijet background estimation does not rely on the simulation as it is obtained directly from a fit to the $m_{\mathrm{jj}}$ distribution. The following parameterization ansatz is adopted to fit the distribution in the $m_{\mathrm{jj}}$ range from $1.1\TeV$ up to the last data point of the inclusive, ``1b'' and ``2b'' mass distributions separately,
\begin{equation}
  f(z)=p_{1}(1-z)^{p_{2}}z^{p_{3}},
  \label{equ:fun}
\end{equation}
where $p_i$ are free parameters and $z=m_{\mathrm{jj}}/\sqrt{s}$.
This ansatz was used in previous searches~\cite{ATLAS:2015nsi} and is found to provide a satisfactory fit to leading-order {\sc Pythia8} multijet MC simulation at $\sqrt{s}=13\TeV$.
Employing Wilks' theorem~\cite{wilks}, a log-likelihood statistic is used to confirm that no additional parameters are needed to model these distributions for a data set as large as the one used for this analysis.

The results of the fits are shown in Figure~\ref{fig:fit}. The fits of this ansatz to the data without considering systematic uncertainties return $p$-values of 0.73, 0.90 and 0.66 for the inclusive, ``1b'' and ``2b'' categories respectively. The $p$-value was calculated as a goodness-of-fit measure using a $\chi^2$ test statistic determined from pseudo-experiments.

The lower panels of Figure~\ref{fig:fit} show the significances of bin-by-bin differences between the data and the fit. These equivalent Gaussian significances are calculated from the Poisson probability, considering only statistical uncertainties.
\begin{figure}[t]
  \center
  \subfloat[At least one $b$-tag]{
 \includegraphics[width=0.45\linewidth]{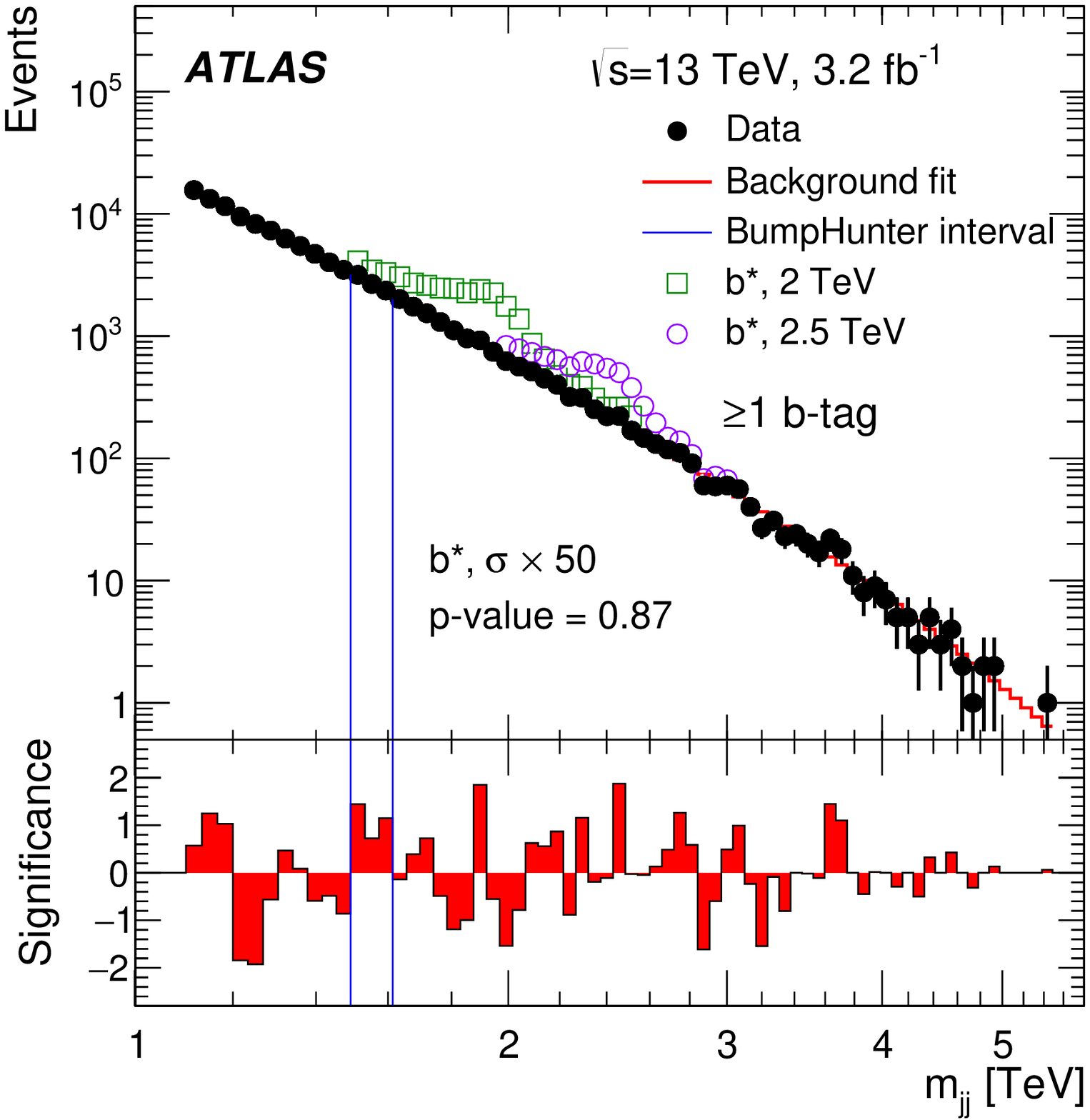}\hfill
  }
  \subfloat[Double $b$-tag]{
  \includegraphics[width=0.45\linewidth]{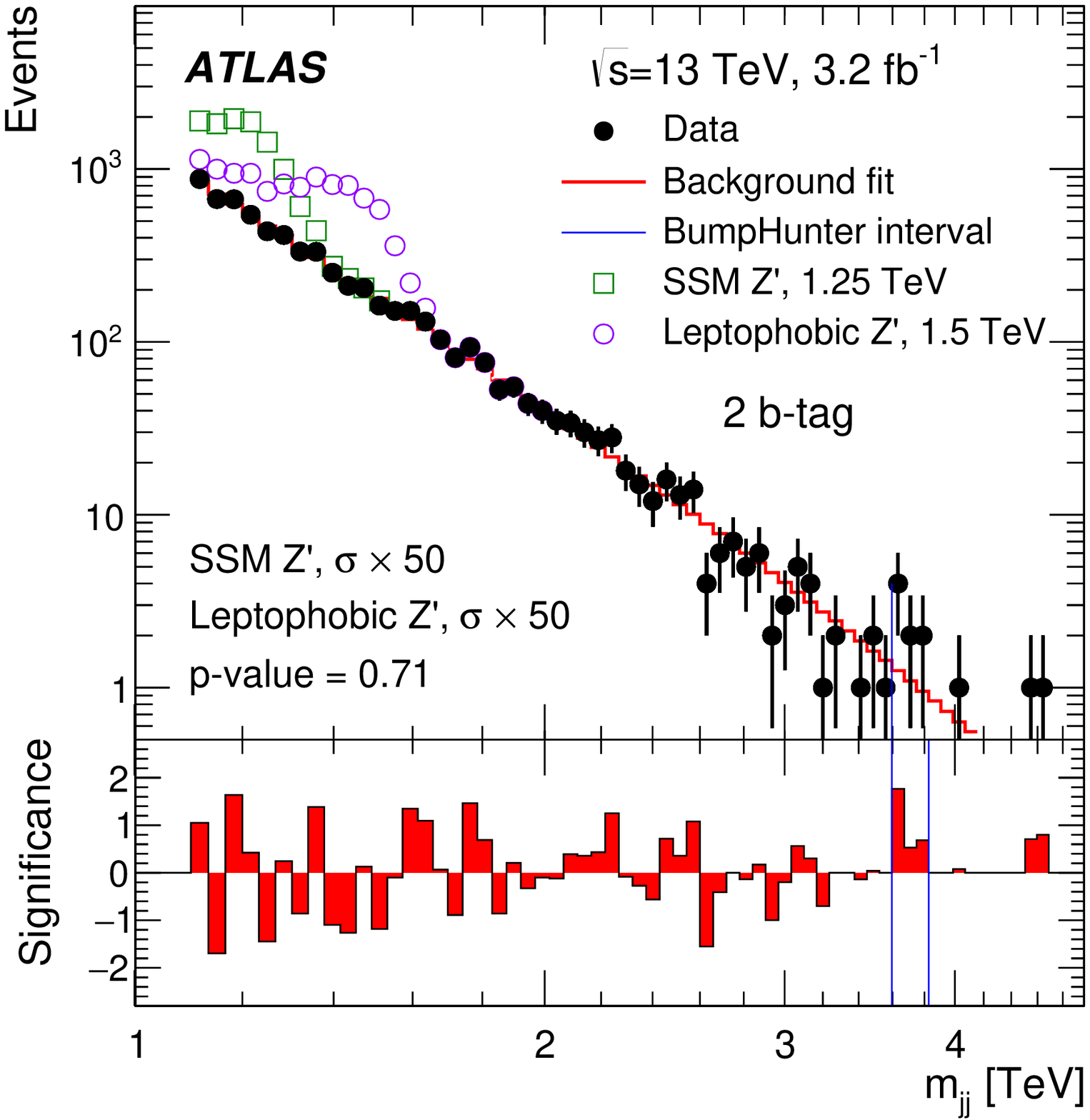}\hfill
  }
  \caption{Dijet mass spectra overlaid with the fits to the background function together with the results from {\sc BumpHunter} and benchmark signals scaled by a factor of 50. The most discrepant region is indicated by the two blue lines. The lower panels show the significances per bin of the data with respect to the background fit, in terms of the number of standard deviations, considering only the statistical fluctuations.
The distributions are shown for the (a) ``1b'' and (b) ``2b'' categories.}
  \label{fig:fit}
\end{figure}

The statistical significance of any localized excess in the dijet mass distribution is quantified using the
{\sc BumpHunter} algorithm~\cite{bumphunter}. The algorithm compares the binned $m_{\mathrm{jj}}$ distribution of the data to the fitted
background estimate, considering contiguous mass intervals in all possible locations, from a width of two bins to one-half of the distribution. For each interval in the scan, it computes the significance of any excess found. The algorithm
identifies the intervals 1493--1614$\GeV$ in the ``1b'' and 3596--3827$\GeV$ in the ``2b'' sample, indicated by the two
vertical lines in Figure~\ref{fig:fit}, as the most discrepant intervals. The statistical significance of these outcomes is
evaluated using the ensemble of Poisson outcomes across all intervals scanned, by applying the algorithm to many pseudo-data
samples drawn randomly from the background fit. Without including systematic uncertainties, the probability that fluctuations of the background model would produce excesses at least as significant as those observed in the data, anywhere in the distribution, is greater than $60\%$ in the ``1b'' and ``2b'' categories. Thus, there is no evidence of localized contributions to the mass distribution from BSM phenomena.

\section{Systematic uncertainties}
\label{sec:syst}

Uncertainties in the parameters of the fitted background function Eq.~\eqref{equ:fun} are evaluated by fitting the ansatz to pseudo-data drawn via Poisson fluctuations around the fitted background model. The uncertainty in the prediction in each $m_{\mathrm{jj}}$ bin is taken to be the root mean square of the function value for 10000 generated pseudo-experiments. To estimate an uncertainty due to
the choice of background parameterization, one additional degree of freedom, $z^{p_{4}\log(z)}$, is appended as a multiplicative factor to the nominal ansatz (Eq.~\eqref{equ:fun}), and the difference between the estimated parameters from the two fits is taken as an uncertainty.

The uncertainty in the jet energy scale is estimated using various methods in $8\TeV$ data, corrected to the new centre-of-mass energy by taking the difference between the $8\TeV$ and $13\TeV$ runs into  account using MC simulation~\cite{ATL-PHYS-PUB-2015-015}.
The jet energy scale uncertainty used in this analysis relies on a set of three nuisance parameters~\cite{ATL-PHYS-PUB-2015-014}.
For untagged jets it is within the range 1--5\% for jet transverse momenta greater than $200\GeV$.

The relative additional uncertainty in the energy scale of $b$-tagged jets is estimated using the MC samples and verified with data following the method described in Ref.~\cite{PERF-2012-01}.
The ratio $r_{\mathrm{trk}}$ of the sum of track transverse momenta inside the jet to the total jet transverse momentum measured in the calorimeter is used for this estimate.
The double ratio of $r_{\mathrm{trk}}$ from data and simulation is formed and compared for inclusive jets and $b$-jets.
The estimated relative additional uncertainty for jets with $200< p_{\mathrm{T}} <800\GeV$ is found to be less than 2.6\%, and this value is subsequently used in the higher $p_{\mathrm{T}}$ regions. This relative uncertainty is applied in addition to the nominal jet energy scale uncertainty.
The maximum uncertainty for $b$-tagged jets is estimated to be 6\% and is conservatively applied to all $p_{\mathrm{T}}$ regions.

The uncertainty in the jet energy resolution is estimated using the same method as the untagged jet energy scale uncertainty and relies on an  additional Gaussian smearing of the reconstructed jet energies in MC simulation.
For jets with $p_\mathrm{T}>50\GeV$, the uncertainty is less than 2\%.

The uncertainty introduced by the application of the $b$-tagging algorithm is the largest systematic uncertainty in the analysis. The uncertainty in the measured tagging efficiency of $b$-jets is estimated by studying $t\bar{t}$ events in $13\TeV$ data for jet $p_\mathrm{T}$ up to $200\GeV$ ~\cite{ATL-PHYS-PUB-2015-039}.
The uncertainties in the measured rate of mistagging $c$-jets and light-flavour jets are estimated in $8\TeV$ data. The uncertainties are extrapolated to $13\TeV$, taking into account the addition of the new IBL system as well as reconstruction and tagging improvements. An additional term is included to extrapolate the measured uncertainties to the high-$p_\mathrm{T}$ region of interest. This term is calculated from simulated events by considering variations on the quantities affecting the $b$-tagging performance such as the impact parameter resolution, percentage of poorly measured tracks, description of the detector material, and track multiplicity per jet. 
The dominant effect on the uncertainty when extrapolating at high-$p_\mathrm{T}$ is related to the different tagging efficiency when smearing the tracks impact parameters based on the resolution measured in data and simulation. The difference in the impact parameter resolution is due to effects from alignment, dead modules and additional material not properly modelled in the simulation. The impact of the $b$-tagging efficiency uncertainty increases with jet $p_\mathrm{T}$ and reaches 50\% above $2\TeV$.

\section{Results}
\label{sec:result}

Due to the absence of a signal, 95\% credibility-level upper limits are set on the cross section for new processes that would produce a contribution to the dijet mass distribution with $b$-tagging. The signal shapes are taken as provided by $b^*\to bg$ and $Z'\to b\bar{b}$ production processes.

The limits on $b^*$ and $Z'$ cross sections are shown in Figures~\ref{fig:limits_bstar}
and \ref{fig:limits_zprime}.
The limits were obtained using a Bayesian method \cite{Aad:2011aj}.
The Bayesian credible intervals were calculated using  a posterior probability density from the likelihood function for the observed mass spectrum  obtained by a fit to the background (Eq.~\eqref{equ:fun}), while the signal shape was derived from MC simulations. The limit is interpolated between discrete values of the mass to create a continuous curve.
The systematic uncertainties associated with the uncertainty in the integrated luminosity, jet energy scale, jet energy resolution, $b$-tagging and alternative fit functions are all included in the limit-setting.

Figure~\ref{fig:limits_bstar} shows that the $b^*$ model, with the decay to $g+b(\bar{b})$, is excluded for $b^*$ masses from $1.1\TeV$ up to $2.1\TeV$ at leading-order in QCD.
\begin{figure}
\begin{center}
\includegraphics[width=0.95\linewidth, angle=0]{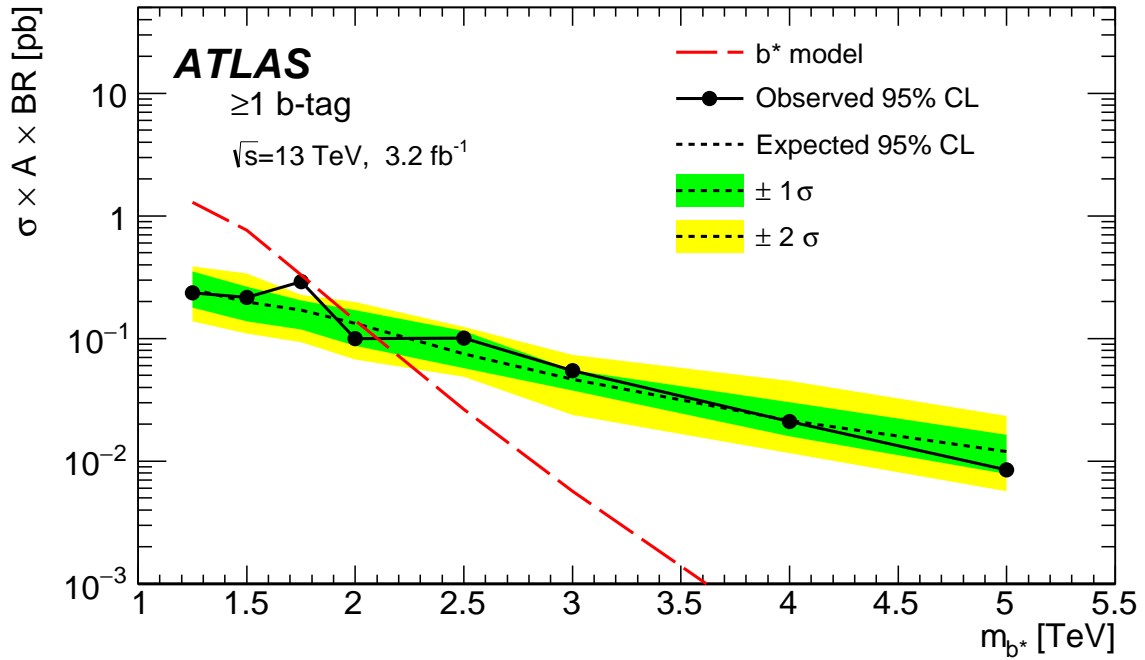}\hfill
\end{center}
\caption{
Observed (filled circles) and expected (dotted line) 95\% credibility-level upper limits on the cross section $\sigma$
times acceptance $A$ times branching ratio $\rm BR$, including kinematic acceptance and $b$-tagging efficiencies, for the $b^*$ model. The dashed line shows the cross section calculated from theory.
The plot shows the case with a single $b$-tag.
}
\label{fig:limits_bstar}
\end{figure}
Figure~\ref{fig:limits_zprime} shows that the leptophobic $Z'\to b\bar{b}$ model with SM-like couplings to quarks is excluded up to 1.5$\TeV$ at leading-order in QCD. The present data are not sufficient to provide an exclusion limit for the SSM $Z'$ model.
\begin{figure}
\begin{center}
\includegraphics[width=0.95\linewidth, angle=0]{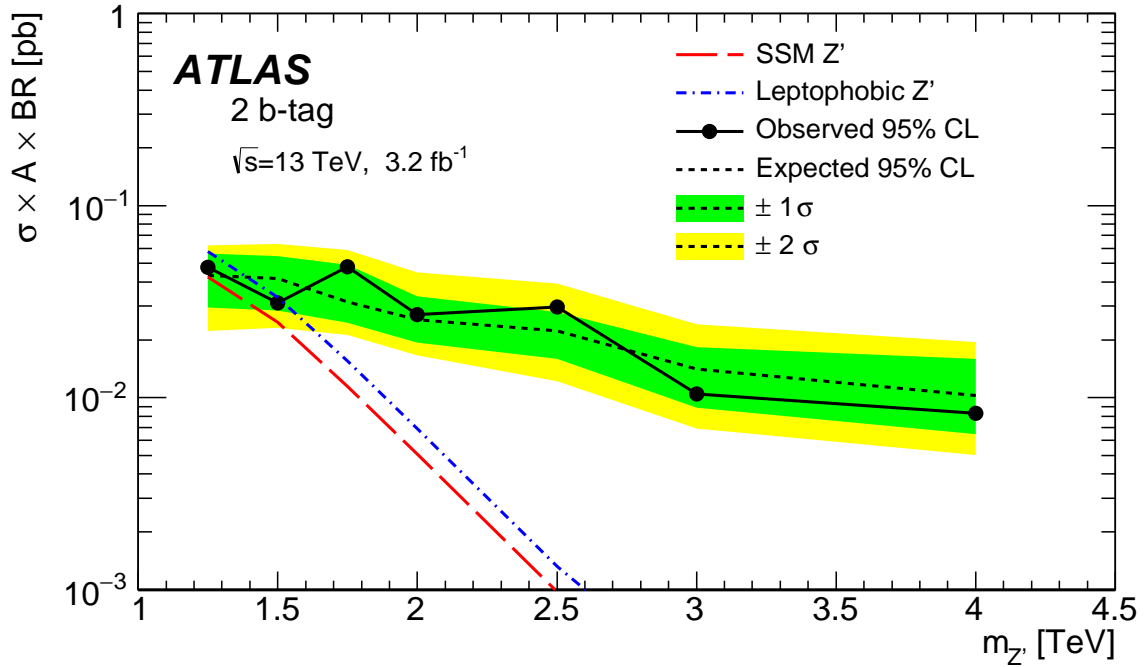}
\end{center}
\caption{
Observed (filled circles) and expected (dotted line) 95\% credibility-level upper limits on the cross section $\sigma$ times the acceptance $A$ times branching ratio $\rm BR$, including kinematic acceptance and $b$-tagging efficiencies, for the $Z'\rightarrow b\bar{b}$ models. The dashed lines show the cross section calculated from theory.
The plot shows the case with double $b$-tag.
}
\label{fig:limits_zprime}
\end{figure}

As shown in Figure~\ref{fig:limits_gauss}, narrow resonance contributions of various widths with visible cross sections $\sigma\times A\times \rm BR$
ranging from approximately 0.4 to 0.001 pb are excluded in the mass range 1.5--5.0$\TeV$.
These limits should be used when long low-mass off-shell tails from PDFs and non-perturbative
effects on the narrow resonance signal shape can be safely truncated or neglected and, after applying the selection described in Section~\ref{sec:selection}, the reconstructed
mass distribution approximates a Gaussian distribution.
For a detailed description of how to use these limits, see the instructions in Ref.~\cite{EXOT-2013-11}.
To estimate the $b$-tagging efficiency, invariant-mass-dependent correction factors as given in Figure~\ref{fig:btageff} can be used.
\begin{figure}
\begin{center}
   \subfloat[At least one $b$-tag] {
\includegraphics[width=0.60\linewidth, angle=0]{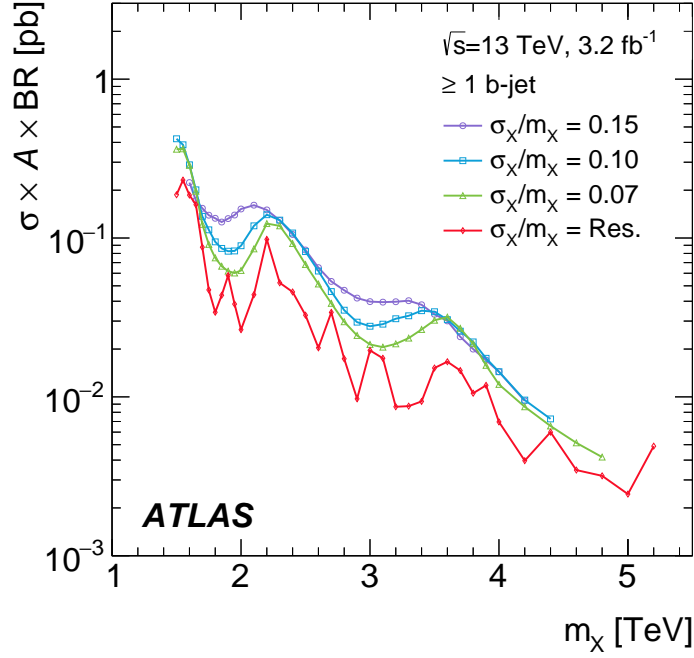}\hfill
   } \\
   \subfloat[Double $b$-tag] {
 \includegraphics[width=0.60\linewidth, angle=0]{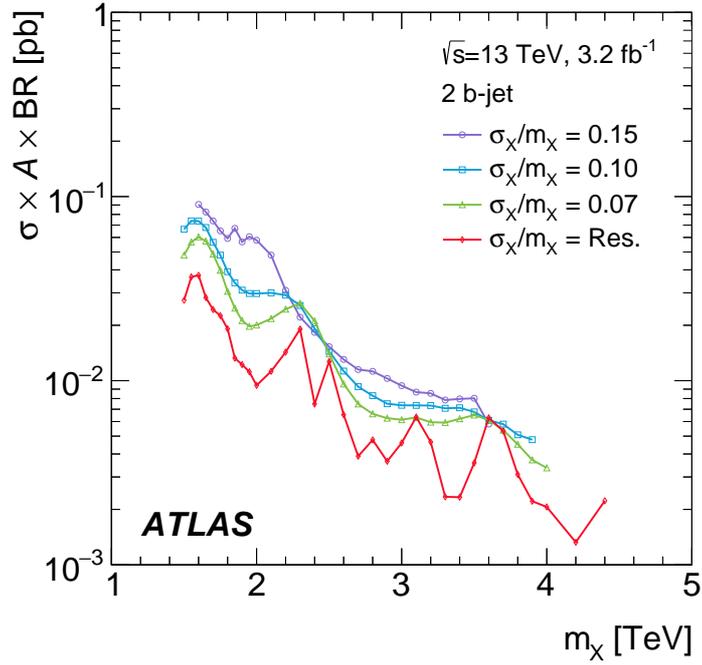}
   }
\end{center}
\caption{
The 95\% credibility-level upper limits on the cross section $\sigma$
times the acceptance $A$ times branching ratio $\rm BR$, including kinematic acceptance and $b$-tagging efficiencies, for resonances exhibiting a generic Gaussian shape.
The circles, squares and triangles correspond to the cases where the width of the Gaussian signal is 15\%, 10\% or 7\% of the signal mass. The figure also shows as a red line the case where the width is given by the dijet mass resolution, which is as low as 2\% at $5.0\TeV$.
The plots show the limits obtained, applying (a) at least one $b$-tag and (b) double $b$-tag.}
\label{fig:limits_gauss}
\end{figure}

\section{Summary}
\label{sec:summary}

A search for new resonances decaying to jets with a single or double $b$-tag in $pp$ collisions with the ATLAS detector at the LHC is presented.
The dataset corresponds to an integrated luminosity of 3.2 \ifb~collected at $\sqrt{s}=13\TeV$ in 2015. The studies use the dijet invariant mass $m_{\mathrm{jj}}$ in the range of 1.1--5.0$\TeV$ with $b$-tagging applied to the leading and sub-leading jets and categorize the events according to their $b$-jet multiplicity.

The background from jets initiated by $b$-quarks is well described by the leading-order parton-shower models. The dijet background is also well described by the analytic fit function with three parameters which is used in the light-flavour dijet analysis \cite{ATLAS:2015nsi}.

No evidence of a significant excess of events is found compared to the expectations of the Standard Model. The largest observed local excess is less than 2$\sigma$ for both the single and double $b$-tag channels.

The expected contribution from the $b^*$ model is excluded in the mass range 1.1--2.1$\TeV$ at leading-order in QCD using the single $b$-jet channel.
The results can not exclude contributions from the SSM $Z'\to b\bar{b}$ model in the mass range 1.1--5.0$\TeV$ in the double $b$-jet channel. For the leptophobic $Z'$ model with SM-like couplings to quarks, the mass range 1.1--1.5$\TeV$ is excluded at leading-order in QCD in this channel.

This analysis excludes generic high-mass particles decaying to two jets, where one or two jets originate from $b$-quarks, with visible cross sections ranging from 0.4 to 0.001 pb in the mass range 1.1--5.0$\TeV$. The exclusion limits are applicable for resonances exhibiting a Gaussian shape and width similar to the $b^*$ or $Z'$ models. The limits were calculated assuming that the width of the Gaussian signal is 15\%, 10\% or 7\% of its mass.

\section*{Acknowledgements}

We thank CERN for the very successful operation of the LHC, as well as the
support staff from our institutions without whom ATLAS could not be
operated efficiently.

We acknowledge the support of ANPCyT, Argentina; YerPhI, Armenia; ARC, Australia; BMWFW and FWF, Austria; ANAS, Azerbaijan; SSTC, Belarus; CNPq and FAPESP, Brazil; NSERC, NRC and CFI, Canada; CERN; CONICYT, Chile; CAS, MOST and NSFC, China; COLCIENCIAS, Colombia; MSMT CR, MPO CR and VSC CR, Czech Republic; DNRF and DNSRC, Denmark; IN2P3-CNRS, CEA-DSM/IRFU, France; GNSF, Georgia; BMBF, HGF, and MPG, Germany; GSRT, Greece; RGC, Hong Kong SAR, China; ISF, I-CORE and Benoziyo Center, Israel; INFN, Italy; MEXT and JSPS, Japan; CNRST, Morocco; FOM and NWO, Netherlands; RCN, Norway; MNiSW and NCN, Poland; FCT, Portugal; MNE/IFA, Romania; MES of Russia and NRC KI, Russian Federation; JINR; MESTD, Serbia; MSSR, Slovakia; ARRS and MIZ\v{S}, Slovenia; DST/NRF, South Africa; MINECO, Spain; SRC and Wallenberg Foundation, Sweden; SERI, SNSF and Cantons of Bern and Geneva, Switzerland; MOST, Taiwan; TAEK, Turkey; STFC, United Kingdom; DOE and NSF, United States of America. In addition, individual groups and members have received support from BCKDF, the Canada Council, CANARIE, CRC, Compute Canada, FQRNT, and the Ontario Innovation Trust, Canada; EPLANET, ERC, FP7, Horizon 2020 and Marie Sk{\l}odowska-Curie Actions, European Union; Investissements d'Avenir Labex and Idex, ANR, R{\'e}gion Auvergne and Fondation Partager le Savoir, France; DFG and AvH Foundation, Germany; Herakleitos, Thales and Aristeia programmes co-financed by EU-ESF and the Greek NSRF; BSF, GIF and Minerva, Israel; BRF, Norway; the Royal Society and Leverhulme Trust, United Kingdom.

The crucial computing support from all WLCG partners is acknowledged
gratefully, in particular from CERN and the ATLAS Tier-1 facilities at
TRIUMF (Canada), NDGF (Denmark, Norway, Sweden), CC-IN2P3 (France),
KIT/GridKA (Germany), INFN-CNAF (Italy), NL-T1 (Netherlands), PIC (Spain),
ASGC (Taiwan), RAL (UK) and BNL (USA) and in the Tier-2 facilities
worldwide.

\newpage

\printbibliography

\newpage

\begin{flushleft}
{\Large The ATLAS Collaboration}

\bigskip

M.~Aaboud$^\textrm{\scriptsize 136d}$,
G.~Aad$^\textrm{\scriptsize 87}$,
B.~Abbott$^\textrm{\scriptsize 114}$,
J.~Abdallah$^\textrm{\scriptsize 65}$,
O.~Abdinov$^\textrm{\scriptsize 12}$,
B.~Abeloos$^\textrm{\scriptsize 118}$,
R.~Aben$^\textrm{\scriptsize 108}$,
O.S.~AbouZeid$^\textrm{\scriptsize 138}$,
N.L.~Abraham$^\textrm{\scriptsize 150}$,
H.~Abramowicz$^\textrm{\scriptsize 154}$
% ...
\bigskip
\\
% ...
$^{12}$ Institute of Physics, Azerbaijan Academy of Sciences, Baku, Azerbaijan\\
% ...
$^{65}$ University of Iowa, Iowa City IA, United States of America\\
% ...
$^{87}$ CPPM, Aix-Marseille Universit{\'e} and CNRS/IN2P3, Marseille, France\\
% ...
$^{108}$ Nikhef National Institute for Subatomic Physics and University of Amsterdam, Amsterdam, Netherlands\\
% ...
$^{114}$ Homer L. Dodge Department of Physics and Astronomy, University of Oklahoma, Norman OK, United States of America\\
% ...
$^{118}$ LAL, Univ. Paris-Sud, CNRS/IN2P3, Universit{\'e} Paris-Saclay, Orsay, France\\
% ...
$^{136}$ $^{(a)}$ Facult{\'e} des Sciences Ain Chock, R{\'e}seau Universitaire de Physique des Hautes Energies - Universit{\'e} Hassan II, Casablanca; $^{(b)}$ Centre National de l'Energie des Sciences Techniques Nucleaires, Rabat; $^{(c)}$ Facult{\'e} des Sciences Semlalia, Universit{\'e} Cadi Ayyad, LPHEA-Marrakech; $^{(d)}$ Facult{\'e} des Sciences, Universit{\'e} Mohamed Premier and LPTPM, Oujda; $^{(e)}$ Facult{\'e} des sciences, Universit{\'e} Mohammed V, Rabat, Morocco\\
% ...
$^{138}$ Santa Cruz Institute for Particle Physics, University of California Santa Cruz, Santa Cruz CA, United States of America\\
% ...
$^{150}$ Department of Physics and Astronomy, University of Sussex, Brighton, United Kingdom\\
% ...
$^{154}$ Raymond and Beverly Sackler School of Physics and Astronomy, Tel Aviv University, Tel Aviv, Israel\\
% ...
$^{a}$ Also at Department of Physics, King's College London, London, United Kingdom\\
$^{b}$ Also at Institute of Physics, Azerbaijan Academy of Sciences, Baku, Azerbaijan\\
$^{c}$ Also at Novosibirsk State University, Novosibirsk, Russia\\
$^{d}$ Also at TRIUMF, Vancouver BC, Canada\\
$^{e}$ Also at Department of Physics {\&} Astronomy, University of Louisville, Louisville, KY, United States of America\\
% ...
\end{flushleft}


\end{document}